\documentclass[showpacs,twocolumn,superscriptaddress]{revtex4-2}

\usepackage{graphicx}
\usepackage{physics}
\usepackage{wrapfig}
\usepackage{changepage}
\usepackage{amsmath,amssymb,amsthm,mathrsfs,amsfonts,dsfont,mathtools}
\usepackage[caption=false]{subfig}
\usepackage{epsfig}
\usepackage{color,xcolor}
\usepackage{adjustbox}
\usepackage{tabularx}
\usepackage{array}
\usepackage{multirow}
\usepackage{tabu}
\usepackage{bm}
\usepackage{enumerate}
\usepackage{hyperref}
\usepackage{newtxtext,newtxmath}
\usepackage{makecell} 
\usepackage{url}
\usepackage{booktabs}
\usepackage{nicefrac} 
\usepackage{makeidx}  
\usepackage{leftindex}
\usepackage{algorithm}
\usepackage{algpseudocode}
\usepackage{cleveref}
\usepackage{mwe}

\usepackage{comment}
\usepackage{xurl}

\begin{document}
\title{Expressivity of deterministic quantum computation with one qubit}

\author{Yujin Kim}
\email{k.yujin2228@yonsei.ac.kr}
\affiliation{Department of Statistics and Data Science, Yonsei University, Seoul 03722, Republic of Korea}
\author{Daniel K. Park}
\email{dkd.park@yonsei.ac.kr}
\affiliation{Department of Statistics and Data Science, Yonsei University, Seoul 03722, Republic of Korea}
\affiliation{Department of Applied Statistics, Yonsei University, Seoul 03722, Republic of Korea}

\begin{abstract}
Deterministic quantum computation with one qubit (DQC1) is of significant theoretical and practical interest due to its computational advantages in certain problems, despite its subuniversality with limited quantum resources. In this work, we introduce parameterized DQC1 as a quantum machine learning model. We demonstrate that the gradient of the measurement outcome of a DQC1 circuit with respect to its gate parameters can be computed directly using the DQC1 protocol. This allows for gradient-based optimization of DQC1 circuits, positioning DQC1 as the sole quantum protocol for both training and inference. We then analyze the expressivity of the parameterized DQC1 circuits, characterizing the set of learnable functions, and show that DQC1-based machine learning (ML) is as powerful as quantum neural networks based on universal computation. Our findings highlight the potential of DQC1 as a practical and versatile platform for ML, capable of rivaling more complex quantum computing models while utilizing simpler quantum resources.
\end{abstract}

\maketitle

\def\one{{\mathchoice {\rm 1\mskip-4mu l} {\rm 1\mskip-4mu l} {\rm\mskip-4.5mu l} {\rm 1\mskip-5mu l}}}

\section{Introduction}
\label{sec:intro}

The ability of machines to learn from data and generalize has become indispensable in modern society. Quantum machine learning (QML) leverages the information processing capabilities of quantum systems to redefine the boundaries of machine learning (ML) and data analysis~\cite{qPCA,PhysRevLett.113.130503_QSVM,schuld2019quantum,RigorousRobustQSpeedUp,10.1038/s43588-022-00311-3}. As the development of universal and fault-tolerant quantum computers remains a long-term prospect, exploring the ML capabilities of less powerful but more realistic quantum devices is of significant importance. In this respect, much effort has been focused on utilizing Noisy Intermediate-Scale Quantum (NISQ) devices~\cite{Preskill2018quantumcomputingin} for ML applications~
\cite{Havlicek2019,benedetti_parameterized_2019,cerezo2020variational,bharti2022noisy,kim2023classical} without requiring quantum error correction. However, less attention has been paid to exploring the capabilities of subuniversal quantum computers for ML tasks, another crucial path toward practical QML.

Supervised learning (SL) is a widely-used technique for data-driven predictions and decision-making. It aims to construct a hypothesis $f(x)$ given a sample dataset
$$
\mathcal{D}=\{ (x_1, y_1), \dots, (x_s, y_s)\} \subseteq \mathbb{R}^d \times \mathbb{R},
$$
where $x_i$ represents a feature vector (independent variable) and $y_i$ is its corresponding outcome (dependent variable). It is common to assume that a pair of data samples $(x_i,y_i)$ is drawn from an unknown joint probability distribution $P_{X,Y}$ (i.e. $(x_i,y_i)\sim P_{X,Y}$). Then, for some error function $L$ quantifying the difference between $f(x)$ and $y$, $f(x)$ must be constructed such that $\mathbb{E}_{(x,y)\sim P_{X,Y} }\lbrack  L(f(x),y)\rbrack$ is minimized. In practice, SL selects $f(x)$ from a family of functions parameterized by $\theta$. Thus, the hypothesis can be expressed as $f(x,\theta)$ where $\theta$ is the optimization variable. 
Expressivity characterizes the complexity of the family of functions the parametric function can generate, and it is a crucial property of an ML model.
In QML, the quantum information processors determine the breadth of function classes.
Previous research has focused on investigating the expressivity within the circuit model of universal quantum computation~\cite{schuld2021effect,PhysRevA.107.062612,PhysRevA.109.042421,Funcke2021dimensional,panadero2023regressionsquantumneuralnetworks}. However, it remains unclear how this scenario changes when quantum computation is constrained to subuniversal models. 

In this work, we analyze the expressivity of the deterministic quantum computation with one qubit (DQC1) model. DQC1 is a subuniversal model of quantum computation where only one quantum bit with non-zero purity can be prepared and measured, while the computation can utilize uniformly random bits. Nevertheless, it can outperform classical computers in solving certain computational problems. Moreover, DQC1 is well-suited for ensemble quantum information processors such as those with spin ensembles and magnetic resonance~\cite{doi:10.1073/pnas.94.5.1634,doi:10.1098/rsta.2011.0352,NMRSimulator,Kyungdeock2015,Lu2016}. Therefore, understanding the ML capabilities of DQC1 in terms of expressivity is crucial for advancing both the theory and practicality of QML. Although previous works have shown that DQC1 can be applied to ML tasks~\cite{PhysRevA.97.032327,DQC1_kernel}, the fundamental question of which function classes DQC1 can generate remains unanswered. To address this, we introduce parameterized DQC1 circuits as an ML model. These circuits incorporate two sets of unitary gates: one for data embedding and the other with learnable parameters for optimization. We demonstrate that the gradient of the measurement outcome of a DQC1 circuit with respect to its parameters can be computed directly using the DQC1 protocol. This allows for gradient-based training of DQC1 circuits, positioning DQC1 as the sole quantum resource for both training and inference.
We then show that the output of a parameterized DQC1 circuit can be represented as a partial Fourier series. The maximum number of orthogonal basis functions grows exponentially with the number of uniformly random bits (i.e., qubits in completely mixed states) and the number of data-embedding layers (thus, the circuit depth). 
Remarkably, this number can match what universal parameterized quantum circuits achieve~\cite{schuld2021effect} by doubling either the number of uniformly random bits or the circuit depth, incurring only a constant overhead.
These findings underscore the potential of DQC1 as a versatile platform for quantum machine learning, offering insights into its capability to rival more complex quantum models while utilizing simpler quantum resources.

The remainder of the paper is organized as follows. Section~\ref{sec:DQC1} provides a brief review of DQC1 to establish the background. In Sec.~\ref{sec:DQC1ML}, we describe the ML model based on parameterized DQC1 circuits and show that the gradient can also be computed using DQC1. Section~\ref{sec:expressivity} analyzes the expressivity of DQC1-based ML models in terms of the number of orthogonal basis functions in the Fourier series representation of their outputs. We provide illustrative examples in Sec.~\ref{sec:examples}. In Sec.~\ref{sec:finite}, we examine the scenario where the DQC1 protocol is provided with a thermal equilibrium state at a finite temperature instead of a completely mixed state, as this more closely resembles realistic physical systems. In Sec.~\ref{sec:multi-measurement}, we explore the machine learning capabilities of an extension of DQC1 that incorporates versions with multiple measurement qubits. Discussions, conclusions, and suggestions for future work are presented in Sec.~\ref{sec:conc}.

\section{DQC1}
\label{sec:DQC1}
This section provides a brief review on DQC1 to setup the background for the main results in this work. DQC1 is a model of quantum computation equipped with a single signal qubit initialized with a non-zero polarization denoted by $\alpha$, along with $n$ uniformly random bits, the capability to apply arbitrary unitary transformations, and the ability to measure the expectation of the Pauli $Z$ observable ($\sigma_z$) on the signal qubit~\cite{DQC1PhysRevLett.81.5672}. It is subuniversal in the sense that only one quantum bit, which is not necessarily pure, can be prepared and measured. The uniformly random bits are typically realized by a quantum system prepared in a maximally mixed state. We refer to this part, which undergoes the controlled unitary operation but is not directly measured, as the working register. Although less powerful compared to universal quantum computers, it is conjectured that DQC1 can solve certain computational problems exponentially faster than classical computers~\cite{DQC1PhysRevLett.81.5672,PhysRevA.72.042316,DQC1complexity}. An outstanding example of this is the problem of estimating the normalized trace of an $n$-qubit unitary operator, $U$, for which the quantum advantage can be achieved if $U$ can be implemented using $O(\text{poly}(n))$ elementary quantum gates.

\begin{figure}[t]
    \centering
\includegraphics[width=0.8\columnwidth]{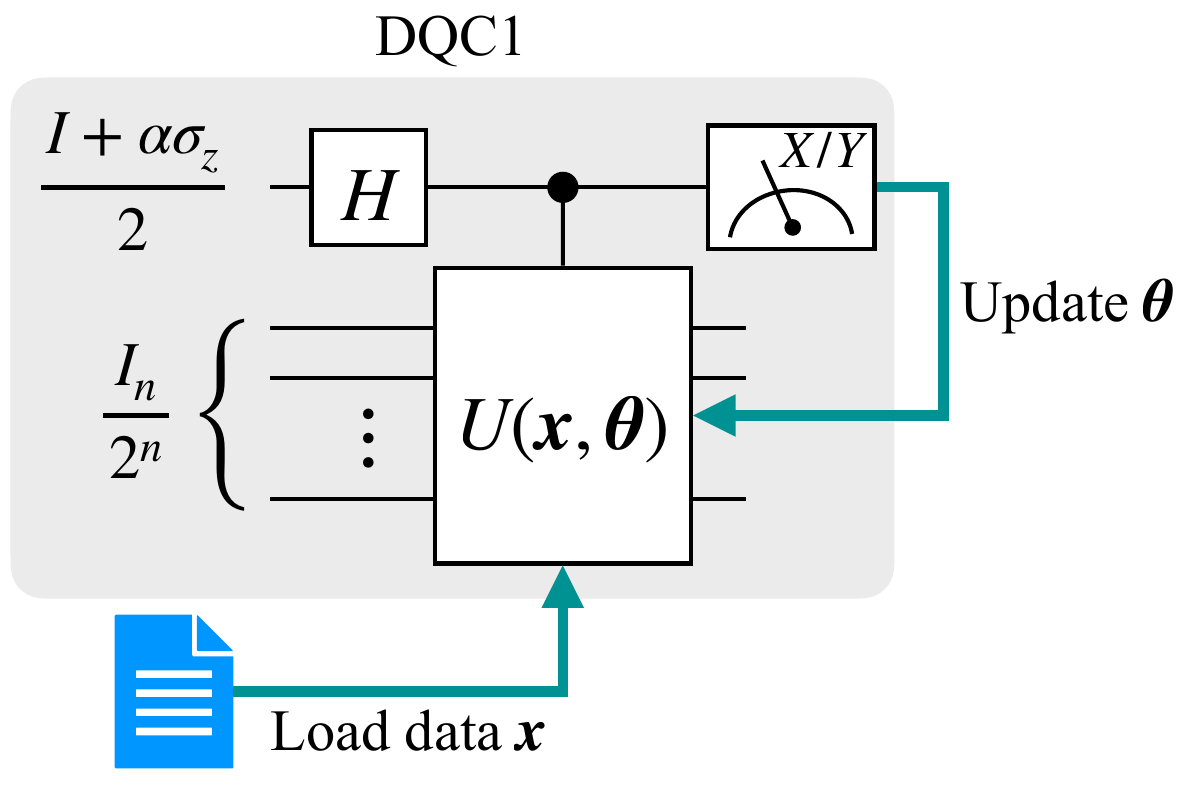}
    \caption{\label{fig:dqc1}Quantum circuit diagram illustrating the DQC1 protocol for estimating the normalized trace of an $n$-qubit unitary operator $U(\boldsymbol{x},\boldsymbol{\theta})$, which is used to construct an ML model. The signal qubit (top qubit) is initialized with non-zero purity ($\alpha > 0$) and measured in either the $\sigma_x$ or $\sigma_y$ basis. The measurement outcome serves to update the parameters $\boldsymbol{\theta}$ of the unitary, facilitating machine learning, with $\boldsymbol{x}$ representing input data.}
\end{figure}

To estimate the trace using DQC1, the following protocol is employed. The process first applies a Hadamard gate to the signal qubit initialized in $(I+\alpha\sigma_z)/2$, where $I$ denotes the $2\times 2$ identity matrix and $\sigma_i$ is the Pauli operator with $i\in\lbrace x,y,z\rbrace$. Then the controlled unitary gate $|0\rangle\langle 0|\otimes I_n+|1\rangle\langle 1|\otimes U$ is applied to the signal qubit and $n$ uniformly random bits, where $I_n$ denotes the $2^n\times 2^n$ identity matrix and the signal qubit acts as the control qubit. This operation prepares the following state:
\begin{equation}\label{eq:measure}
\rho=\frac{1}{2^{n+1}}\left(I_{n+1}+\alpha\left(|0\rangle\langle 1|\otimes U^{\dagger}+|1\rangle\langle 0|\otimes U\right)\right).
\end{equation}
The protocol concludes by measuring the expectation values of Pauli $X$ and $Y$ observables ($\sigma_x$ and $\sigma_y$) on the signal qubit, resulting in
\begin{equation}
\label{eq:observables}
\langle \sigma_x\rangle=\frac{\alpha}{2^n}\mathrm{Re}\left(\mathrm{tr}\left(U\right)\right),\quad \langle \sigma_y\rangle=\frac{\alpha}{2^n}\mathrm{Im}\left(\mathrm{tr}\left(U\right)\right).
\end{equation}
Given the ability of measuring $\langle \sigma_z\rangle$ and applying arbitrary single-qubit gates, $\langle\sigma_x\rangle$ and $\langle\sigma_y\rangle$ can be measured via Clifford transformations $H\sigma_z H=\sigma_x$ and $S\sigma_xS^{\dagger}=\sigma_y$. Repeating the DQC1 circuit $O(\log(1/\delta)/(\alpha\epsilon)^2)$ times facilitates the estimation of the expectation values within $\epsilon$ with a probability of error $\delta$ \cite{estimate}. Figure ~\ref{fig:dqc1} illustrates the quantum circuit for the DQC1 protocol within the gray box. The unitary operator in the figure is parameterized by $\boldsymbol{x}$ and $\boldsymbol{\theta}$, where $\boldsymbol{\theta}$ is updated based on the measurement outcomes for a specific machine learning task. Further details on this process will be provided in the following section.

\section{DQC1 for ML}
\label{sec:DQC1ML}
Let us denote the set of functions that be expressed by a DQC1 protocol as $\mathcal{F}$ and an element of the set as $f(\boldsymbol{x},\boldsymbol{\theta})$. Without loss of generality, we set $\alpha=1$ henceforth, as it can be efficiently polarized via various algorithmic cooling techniques~\cite{boykin2002algorithmic,schulman1999molecular,Park2016,elias2011semioptimal,lin2024thermodynamic,HBAC_daniel}. In general, a DQC1-based ML model with $n$ uniformly random bits can be defined as
\begin{equation}
\label{eq:fx}    
f(\boldsymbol{x},\boldsymbol{\theta}) \in \mathcal{F} = \left\lbrace \frac{1}{2^n}\mathrm{tr}\left(U(\boldsymbol{x},\boldsymbol{\theta})\right) : U^{\dagger}=U^{-1}, \boldsymbol{\theta}\in \boldsymbol{\Theta} \right\rbrace ,
\end{equation}
where $\boldsymbol{x}$ is the input data and $\boldsymbol{\Theta}$ denotes the real-valued parameter space. Specifically, we consider the unitary operator in the form of 
\begin{equation}\label{eq:U}
U(\boldsymbol{x},\boldsymbol{\theta})=\prod_{l=1}^{L}W_l(\boldsymbol{\theta}_l)V_l(\boldsymbol{x}_{l}).    
\end{equation}
In this expression, $\boldsymbol{\theta}_l$ and $\boldsymbol{x}_l$ represent subsets of the parameter vector $\boldsymbol{\theta}$ and the data vector $\boldsymbol{x}$, respectively, which are loaded at the $l$th step. The trainable unitary $W_l(\boldsymbol{\theta}_l)$ can be written as 
\begin{equation}
W_l(\boldsymbol{\theta}_l)=\prod_{k=1}^{k'}\exp(-i (\boldsymbol{\theta}_l)_k H_{lk})T_{lk},
\end{equation}
where $(\boldsymbol{\theta}_l)_k$ represents the $k$th element of $\boldsymbol{\theta}_l$, $H_{lk}$ is an $2^n\times 2^n$ Hermitian operator and $T_{lk}$ is an unparametrized unitary. This form of $U(\boldsymbol{x},\boldsymbol{\theta})$ is a reasonable choice since an $n$-qubit Pauli operator, $\sigma_k\in\mathcal{P}_n=\lbrace I,\sigma_x,\sigma_y,\sigma_z\rbrace^{\otimes n}$, commutes with itself and $\lbrace \exp(-i (\boldsymbol{\theta}_l)_k \sigma_{k}): \sigma_k\in\mathcal{P}_n\rbrace$ forms a universal gate set.

The partial derivative of the model function with respect to a parameter can be computed as
\begin{widetext}
\begin{equation}\label{eq: gradient}
  \frac{\partial f(\boldsymbol{x},\boldsymbol{\theta})}{\partial (\boldsymbol{\theta}_l)_k} = \frac{-i}{2^n}\mathrm{tr}\left(H_{lk}\left(\prod_{j=k}^{k'}\exp(-i(\boldsymbol{\theta}_l)_jH_{lj})T_{lj}\right)V_l(\boldsymbol{x}_{l})\left(\prod_{j=l+1}^{L}W_{j}(\boldsymbol{\theta}_j)V_j(\boldsymbol{x}_{j})\right)\left(\prod_{j=1}^{l-1}W_{j}(\boldsymbol{\theta}_j)V_j(\boldsymbol{x}_{j})\right)\left(\prod_{j=1}^{k-1}\exp(-i(\boldsymbol{\theta}_l)_jH_{lj}) T_{lj}\right) \right).
\end{equation}
\end{widetext}
Therefore, if $H_{lk}$ is a unitary (e.g. $H_{lk}\in \mathcal{P}_n$), the gradient of $f(\boldsymbol{x},\boldsymbol{\theta})$ with respect to $\boldsymbol{\theta}$ can be obtained via DQC1, and the gradient-based optimization techniques can be employed for training the model. In the special case where the trainable unitary is in the form of 
\begin{equation}
    W_l(\boldsymbol{\theta}_l)=\exp(-i \sum_{k=1}(\boldsymbol{\theta}_l)_k \sigma_k),
\end{equation}
and all $\sigma_k$ operators in the summation commute with each other, the partial derivative in Eq.~(\ref{eq: gradient}) can be further simplified to
\begin{equation}
    \frac{\partial f(\boldsymbol{x},\boldsymbol{\theta})}{\partial (\boldsymbol{\theta}_l)_k} = \frac{-i}{2^n}\mathrm{tr}\left(\sigma_k \prod_{j=l}^{L}W_{j}(\boldsymbol{\theta}_{j})V_{j}(\boldsymbol{x})\prod_{j=1}^{l-1} W_{j}(\boldsymbol{\theta}_{j})V_{j}(\boldsymbol{x}) \right).
\end{equation}

The ability of the DQC1 protocol to estimate the gradient of the ML model with respect to its trainable parameters opens up opportunities for integrating DQC1 with other variational quantum circuits or classical machine learning models~\cite{broughton2020tensorflow,co-design}. For instance, a DQC1 circuit can be combined with classical neural networks for tasks such as transfer learning~\cite{Mari2020transferlearningin,kim2023classical}, self-supervised learning~\cite{QSSL}, few-shot learning~\cite{9951229}, and quantum embedding optimization~\cite{PhysRevA.110.022411}. Such hybrid approaches could enhance the adaptability and performance of the model~\cite{PhysRevA.109.042421}, allowing it to tackle more complex problems.

\section{Expressivity of DQC1}
\label{sec:expressivity}
Given $U(\boldsymbol{x},\boldsymbol{\theta})$ in the form of Eq.~\eqref{eq:U}, the data embedding unitary $V_{l}(\boldsymbol{x}_{l})$ can be constructed by single-qubit rotation gates $V_{l}(\boldsymbol{x}_{l})=\prod_{q=1}^{n}\exp(-i (\boldsymbol{x}_l)_q\sigma_k^{(q)}/2)$, where $\sigma_k^{(q)}$ is a Pauli operator acting on the $q$th qubit and $(\boldsymbol{x}_l)_q$ represents the $q$th component of $\boldsymbol{x}_l$. This form is appropriate as any fixed gates involved in data embedding can be absorbed to $W$ terms before and after $V_l$. The data-embedding circuit can be further simplified to $V_{l}(\boldsymbol{x}_{l})=\prod_{q=1}^{n}\exp(-i (\boldsymbol{x}_l)_q\sigma_z^{(q)}/2)$ by absorbing the gates for diagonalizing the Pauli operator to the $W$ terms. This simplification yields the expression for $V_{l}(\boldsymbol{x}_{l})$ as a diagonal matrix as follows:
\begin{equation}
    V_{l}(\boldsymbol{x}_{l})
    =\exp\,\left(-\frac{i}{2}\sum^{n}_{q=1}
    (\boldsymbol{x}_l)_q\sigma_z^{(q)}\right)=\exp\left(\,i D_l\,\right)
\end{equation}
where $D_l\equiv-
(1/2)\sum_{q=1}^{n} 
(\boldsymbol{x}_i)_q\,\sigma_z^{(q)}$. $D_l$ possesses $2^n$ eigenvalues composed of all possible combinations of sums with $\pm (\boldsymbol{x}_l)_q$, up to a constant.

Now, denoting the element of a matrix at the $i$th row and the $j$th column as $(\cdot)_{ij}$, the trace of the parameterized unitary operator can be expressed as 
\begin{align}
\begin{split}
\label{eq:trace_full}
\mathrm{tr}(U(\boldsymbol{x},\boldsymbol{\theta}))=&\sum^{2^n}_{i=1}\Bigl(W_1(\boldsymbol{\theta}_1)V_1(\boldsymbol{x}_{1})
\cdots
W_L(\boldsymbol{\theta}_L)V_L(\boldsymbol{x}_{L})
\Bigr)_{ii}\\
=&\sum^{2^n}_{i=1}\sum^{2^n}_{k_1=1}\cdots
\sum^{2^n}_{k_L=1}
W_1(\boldsymbol{\theta}_1)_{ik_1}
V_1(\boldsymbol{x}_{1})_{k_1k_1}
\cdots\\
&\qquad\qquad\qquad\times
W_L(\boldsymbol{\theta}_L)_{k_{L-1}k_L}
V_L(\boldsymbol{x}_{L})_{k_Li}
\end{split}    
\end{align}
where $i=k_L$ since $V_l(\boldsymbol{x}_{l})$ is a diagonal matrix for all $l=1,2,\cdots,L)$. This expression shows that the trace can be represented as a linear combination of all possible products of eigenvalues, where each product includes one eigenvalue from $V_l(\boldsymbol{x}_{l})$ for each $l$. Each eigenvalue of $V_l(\boldsymbol{x}_{l})$ is a diagonal element of $\exp(iD_l)$, and thus the product of the eigenvalues can be expressed as $\exp(i\sum_{l=1}^{L}(D_{l})_{k_{l}k_{l}})$, where $k_l \in \lbrace 1,2,3,\ldots,2^n\rbrace$. The  coefficients in the linear combination of the products are determined by the trainable unitaries.

Combining these insights, Eq.~(\ref{eq:trace_full}) can be interpreted as a Fourier-type sum:
\begin{equation}
\label{eq:trace_fourier}
\mathrm{tr}(U(\boldsymbol{x},\boldsymbol{\theta})) = \sum_{\omega \in \Omega} c_{\omega}(\boldsymbol{\theta}) \exp(\,i \omega(\boldsymbol{x})),
\end{equation}
where $\omega=\omega(\boldsymbol{x})\in\mathbb{R}$ is an element of the frequency spectrum $\Omega$ determined by the data-embedding part of $U(\boldsymbol{x},\boldsymbol{\theta})$ and $c_{\omega}(\boldsymbol{\theta})$ is the corresponding coefficient controlled by the trainable unitary part. Specifically,
\begin{equation}
\label{eq:modes}
\exp(\,i\omega(\boldsymbol{x}))
=\exp\left(\,i\sum^{L}_{i=1} (D_{i})_{k_{i}k_{i}}
\right),
\end{equation}
and
\begin{equation}
\label{eq:coeff}
    c_{\omega}(\boldsymbol{\theta})
=(W_{1}(\boldsymbol{\theta}_{1}))_{k_{L}k_{1}}
(W_{2}(\boldsymbol{\theta}_{2}))_{k_{1}k_{2}} \cdots
(W_{L}(\boldsymbol{\theta}_{L}))_{k_{L-1}k_{L}}.
\end{equation}

The cardinality of frequency spectrum $\Omega$ is
\begin{equation}
\label{eq:cardinality}
|\Omega|\leq 2^{nL},
\end{equation}
where the equality is satisfied 
when all $V_{l}(\boldsymbol{x}_{l})$ have different eigenvalues and they produce all different frequencies.
This is the number of orthogonal basis functions of a Fourier series, indicating the degrees of complexity of the quantum model.
Equation~\eqref{eq:cardinality} implies that scaling up the number of qubits (parallel) yields the same level of complexity as increasing the circuit depth (serial).

In comparison, the quantum neural network (QNN)~\cite{10.1038/s41467-018-07090-4} based on universal computation with $n$ qubits yields the following set of functions: 
\begin{equation}\label{eq: qnn model}
f_{\mathrm{u}}(\boldsymbol{x},\boldsymbol{\theta})\in\left\lbrace \langle \boldsymbol{0}_n | U^{\dagger}(\boldsymbol{x},\boldsymbol{\theta})MU(\boldsymbol{x},\boldsymbol{\theta})|\boldsymbol{0}_n\rangle : M^{\dagger}=M \right\rbrace,    
\end{equation}
where $|\boldsymbol{0}_n\rangle = |0\rangle^{\otimes n}$ and $U(\boldsymbol{x},\boldsymbol{\theta})$ takes the same form as in Eq.~(\ref{eq:U}). The cardinality of the frequency spectrum produced by this model is $|\Omega|\le 2^{2n(L-1)}$~\cite{schuld2021effect}. This shows that DQC1 can generate as many orthogonal Fourier basis functions as the universal model simply by increasing the number of qubits or the circuit depth by about a factor of two.

We note that the ability of a single-qubit circuit to approximate a function in the asymptotic limit was also demonstrated in Ref.~\cite{PhysRevA.104.012405}. However, the theorems and proofs presented therein are existence results and are limited to univariate functions. While it is possible to extend this single-qubit approach to multivariate functions by increasing the circuit depth as the number of input variables increases, the expressivity of such an extended model remains to be investigated. Interestingly, the quantum circuit structure in that work can be understood as a special construction within the framework of quantum signal processing (QSP) that implements a broad class of complex polynomial functions~\cite{Low_2017,Martyn_2021}. In this view, the data-embedding unitary acts as the signal operator, the parameterized ansatz serves as the signal processing operator, and the signal basis is parameterized. In contrast, our results demonstrate that multivariate functions can be approximated within the single-qubit framework by utilizing the DQC1 model, for which we provide an explicit analysis of expressivity. Our method entails a trade-off between circuit depth and the number of maximally mixed qubits (or uniformly random bits), allowing a DQC1 circuit to approximate multivariate functions even with a constant circuit depth. Furthermore, by increasing the circuit depth by only a factor of two, the DQC1 model can be made as expressive as the universal quantum circuit model.

\section{Machine Learning Examples}
\label{sec:examples}

\subsection{Function Approximation}

\begin{figure*}[t]
    \centering
    \includegraphics[width=1\textwidth]{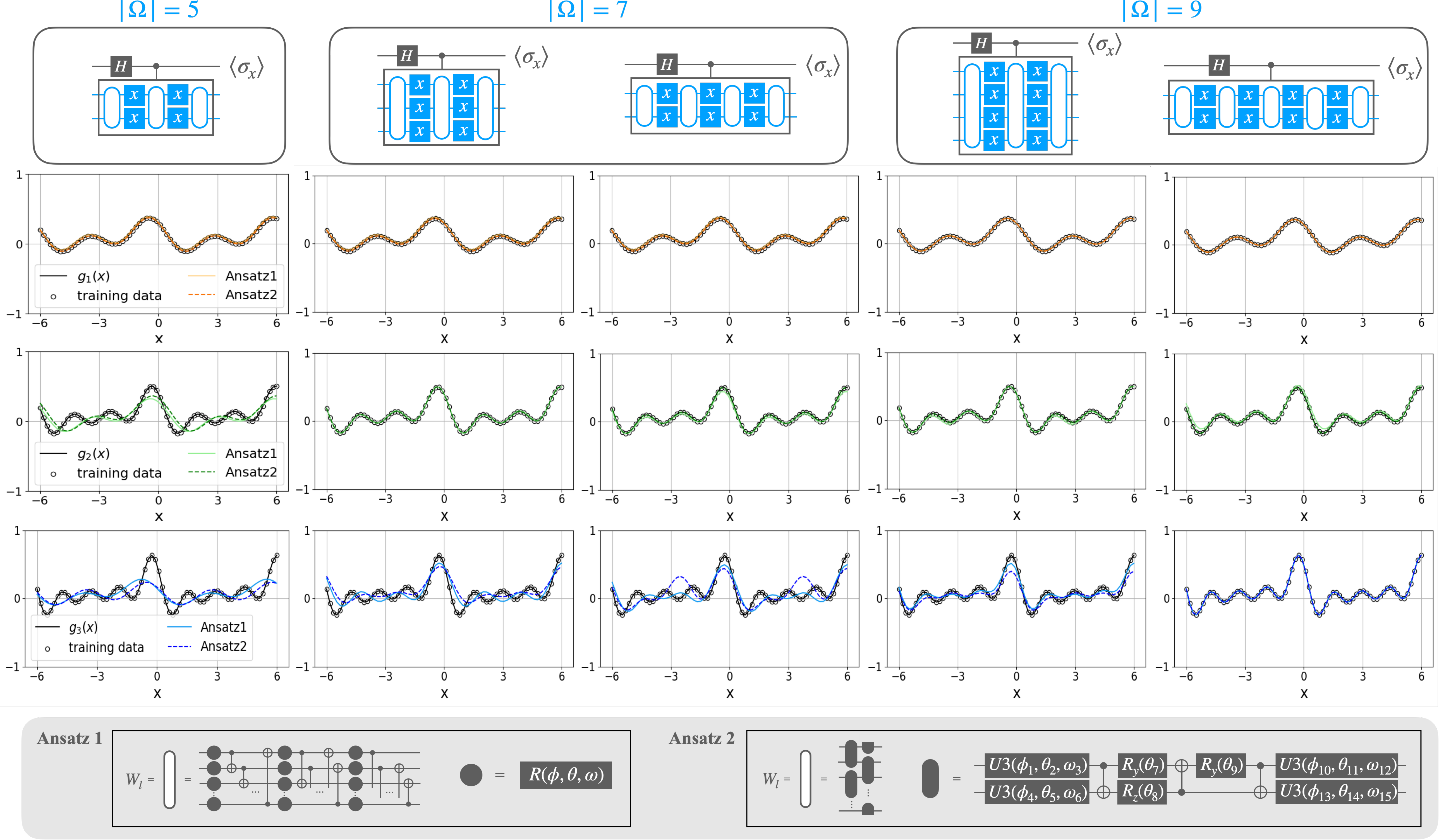}
    \caption{\label{fig:ml example}
    Simulation results of learning functions using DQC1-based ML models $f(\boldsymbol{x},\boldsymbol{\theta})$ with $|\Omega|=5,7,9$. The target functions are the real parts of
    $g_{1}(x)=\sum_{k=-2}^{2}c_{k}\exp(ikx)$, $g_{2}(x)=\sum_{k=-3}^{3}c_{k}\exp(ikx)$, and $g_{3}(x)=\sum_{k=-4}^{4}c_{k}\exp(ikx)$ with $c_0=0.1$ and $c_{k}=0.05+0.05i$ for $k\neq0$. These functions consist of 5, 7 and 9 frequencies, respectively. The two ans\"{a}tze used for the training unitary $W_{l}(\boldsymbol{\theta})$ are depicted at the bottom of the figure. In Ansatz 1, the $R(\phi, \theta, \omega)$ gate denotes an arbitrary single-qubit rotation gate, while in Ansatz 2, the $U3$ gate represents a generic single-qubit rotation gate with three Euler angles, as defined in the main text. The $R_j(\theta)$ gate represents the single-qubit rotation gate around the $j$-axis of the Bloch sphere.}
\end{figure*}

We validate the main findings of the previous section through numerical simulations. In this example, the goal is to learn a univariate function $g(x)$ by training the DQC1-based ML model. 
We set our model as $\mathrm{Re}(f(\boldsymbol{x},\boldsymbol{\theta}))=\mathrm{Re}(\mathrm{tr}(U(\boldsymbol{x},\boldsymbol{\theta}))/2^n)$ by measuring the expectation values of Pauli $X$ $(\sigma_x)$ observable on the signal qubit, as described in Eq.~\eqref{eq:observables}.
The target functions are the real parts of $g_{1}(x)=\sum_{-2}^{2}c_{k}\exp(ikx)$, $
g_{2}(x)=\sum_{-3}^{3}c_{k}\exp(ikx)$, and $g_{3}(x)=\sum_{-4}^{4}c_{k}\exp(ikx)$ with $c_0=0.1$ and $c_{k}=0.05+0.05\,i$ for $k\neq 0$. These functions consist of 5, 7, and 9 frequencies, respectively. The functions with varying frequency spectra are chosen to demonstrate how the size of DQC1 circuits increases with the number of frequencies in the Fourier representation of the target functions.

The univariate input $x$ is encoded using single-qubit $\sigma_x$ rotation gates $V_{l}(x)$, with $V_{L+1}=I_{n}$. Consequently, the diagonal matrix $D_l\,(l\in\lbrace 1,\cdots,L\rbrace)$ has $n+1$ unique eigenvalues corresponding to $-n/2, -(n-2)/2,\cdots, n/2$. An $(L+1)$-layer circuit subsequently generates $(nL+1)$ unique frequencies, 
$-nL/2, -(nL-2)/2,\cdots, nL/2$, satisfying 
\begin{equation}\label{eq:ex cardinality}
    |\Omega|=nL+1.
\end{equation}
This simple encoding scheme is chosen for illustrative purposes. However, an encoding scheme that generates exponentially many frequencies~\cite{PhysRevA.107.012422} can also be utilized in DQC1 models.

Two types of trainable unitaries $W_l(\boldsymbol{\theta_l})$ are chosen for this demonstration, as depicted at the bottom of Fig.~\ref{fig:ml example}.
The first ansatz (Ansatz 1) consists of three layers of a generic strongly entangling circuit~\cite{PhysRevA.101.032308}. In each layer, all qubits are parameterized by an arbitrary single-qubit rotation gate $R(\phi,\theta,\omega)=R_z(\omega)R_y(\theta)R_z(\phi)$ and entangled via CNOT gates in different configurations per layer, totaling $3\times3\times n$ parameters. Here, $R_j(\theta)$ refers to the standard single-qubit rotation gate around the $j$-axis of the Bloch sphere. The second ansatz (Ansatz 2) comprises two layers of nearest-neighbor two-qubit gates using an arbitrary $SU(4)$ circuit as introduced in Ref.~\cite{hur2022quantum}. This circuit is composed of $U3$, $R_y$, $R_z$, and CNOT gates, where $U3(\phi,\theta,\omega)=\exp(i(\theta+\omega)/2)R_z(\theta+\omega)R(\omega,\phi,-\omega)$ is an arbitrary single-qubit gate with three Euler angles. This configuration has $15\times n$ parameters when the number of qubits is even, and $15\times (n-1)$ parameters otherwise. 

For each ansatz, we utilize five DQC1 models, varying the number of qubits $n$ and the number of layers $L$, as illustrated in the top row of Fig.~\ref{fig:ml example}. The cardinality of models corresponds to 5, 7, and 9, as per Eq.~(\ref{eq:ex cardinality}). We verify how well these models can learn target functions with 5, 7, and 9 frequencies, comparing their performance with the theoretical expressivity of the models.

In simulations, 70 sample data points, equally spaced over the interval $x\in[-6,6]$ are used as the training set, with a batch size of 25. Parameter updates are iterated 200 times using the Adam optimizer~\cite{kingma2017adam} with a learning rate of 0.15, both selected based on comparative studies presented in Appendix~\ref{Append: A}, and implemented in Pennylane~\cite{bergholm2020pennylane}. 

Figure \ref{fig:ml example} summarizes the simulation results. 
In each plot, the target function is represented by a black line, and the training data points are shown as open circles. The colored solid lines indicate the training results using ansatz 1, while the colored dotted lines represent the results using ansatz 2.
The training results for $g_1(x)=\sum^{2}_{k=-2}c_k\exp(ikx)$ are shown in the first row of plots. These results demonstrate that all models can fully express the target function since they contain $|\Omega|\geq5$, which is greater than the number of frequencies in $g_1(x)$.
In the second row, the results for $g_2(x)=\sum^{3}_{k=-3}c_k\exp(ikx)$, which has 7 frequencies, show that the model with $|\Omega|=5$ cannot adequately represent the target function. However, the models with $|\Omega|\geq7$ can do so. For $g_3(x)=\sum^{4}_{k=-4}c_k\exp(ikx)$ in the third row, only the models containing $|\Omega|=9$ can match the target function.

The simulation results confirm that DQC1-based ML models achieve sufficient expressive power to learn the target function, provided the circuit employs the number of qubits and data-encoding layers specified by Eq.~(\ref{eq:cardinality}).

As the number of qubits increases, training DQC1 models to accurately predict values that significantly deviate from zero becomes increasingly challenging. This issue is particularly evident in the plot for $g_3(x)$ using four qubits and the three-layer model depicted in the fourth column. We speculate that this phenomenon is related to exponential concentration, a well-known problem in both QNN~\cite{10.1038/s41467-018-07090-4,larocca2024review} and quantum kernel methods~\cite{Thanasilp2024}. This issue will be discussed in more detail in Sec.~\ref{sec:conc}.

\begin{figure}[t]
    \centering
\includegraphics[width=0.9\columnwidth]{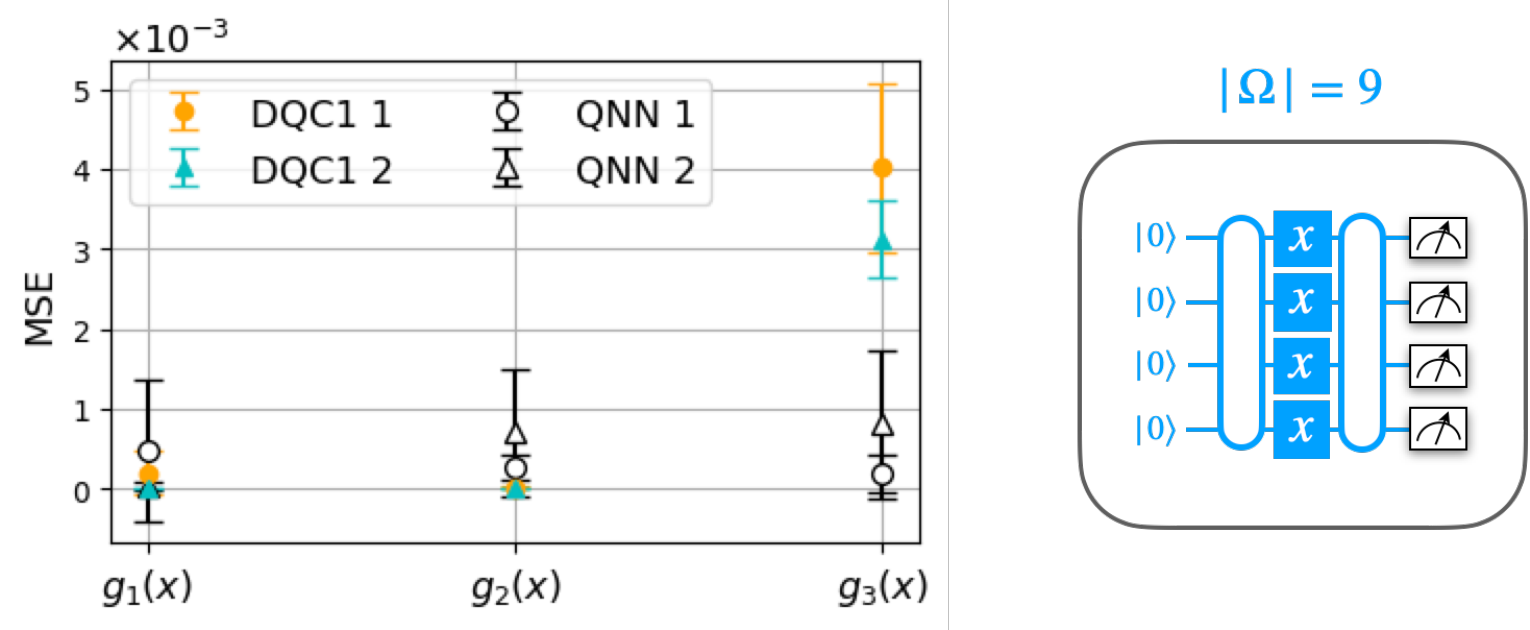}
    \caption{\label{fig:dqc1_qnn} Performance comparison between the DQC1-based ML model and the QNN, evaluated by the MSE of the final optimized models. The markers indicate the mean, and error bars represent standard deviation, calculated from five repetitions with randomly initialized parameters. Circles indicate results using Ansatz 1, and triangles indicate results using Ansatz 2. The circuit used for the QNN is displayed on the right.}
\end{figure}

To assess the effectiveness of the DQC1-based model, we compare its ability to learn functions to that of QNNs.
Figure~\ref{fig:dqc1_qnn} compares the mean squared error (MSE) for the final trained models on the target functions $g_1(x), g_2(x),$ and $g_3(x)$ using circuits with the same expressivity. For the DQC1-based ML models, we selected the 4-qubit circuit with expressivity $|\Omega|=9$, which is the most expressive and shallowest circuit from Fig.~\ref{fig:ml example}. For the QNN models described in Eq.~\eqref{eq: qnn model}, we used a $4$-qubit circuit with one encoding layer, ensuring its expressivity matches that of the DQC1 model (i.e. $|\Omega|=9$). In the QNN models, the measurement operator $M$ is a multi-qubit observable $\sigma_z\otimes \sigma_z\otimes \sigma_z\otimes \sigma_z$, chosen to reflect the distinction from DQC1, which is restricted to single-qubit measurements. For both models, we employed the two ans\"{a}tze shown in Fig.~\ref{fig:ml example}, and identical simulation settings were used including the input data, type of optimizer, and learning rate. The MSE is calculated as $\sum^{N}_{i=1}\,(y_i-f(x_i,\boldsymbol{\theta}_{\mathrm{opt}}))^2/(2N)$, where $y_i$ is the true label for $x_i$ and $\boldsymbol{\theta}_{\mathrm{opt}}$ represents the optimized final parameters for the quantum model $f$. 

The results show that for $g_1(x)$ and $g_2(x)$ (with $\vert \Omega\vert= 5$ and $7$, respectively), the DQC1-based model and QNN achieve similar levels of MSE, with the former performing slightly better. However, for $g_3(x)$, the QNN model achieves lower MSE, which can be attributed to the exponential concentration of the DQC1 model, as mentioned earlier.

\subsection{Classification}

We evaluate the practicality of the DQC1-based ML model for real-world applications through additional experiments, benchmarking its binary classification performance against the QNN model on three distinct datasets.

Specifically, we used the first and second classes of the Iris dataset~\cite{https://doi.org/10.1111/j.1469-1809.1936.tb02137.x}, the 0 and 1 classes from the MNIST image dataset~\cite{lecun2010mnist}, and the first and second classes of the Wine dataset~\cite{Lichman2013}. For each dataset, we selected 100, 400, and 130 data points, respectively, and split them into training and test sets with an 80-20 ratio. For the MNIST and Wine datasets, principal component analysis (PCA) was applied to extract four features, while all datasets subsequently underwent min-max scaling. In all simulations, we used the Adam optimizer with the learning rates of 0.09, 0.005, and 0.004 for the Iris, MNIST, and Wine datasets, respectively, selected through multiple learning rate comparisons. Additionally, the results were obtained by running 40, 70, and 110 epochs for each case.

\begin{figure}[t]
    \centering
\includegraphics[width=\columnwidth]{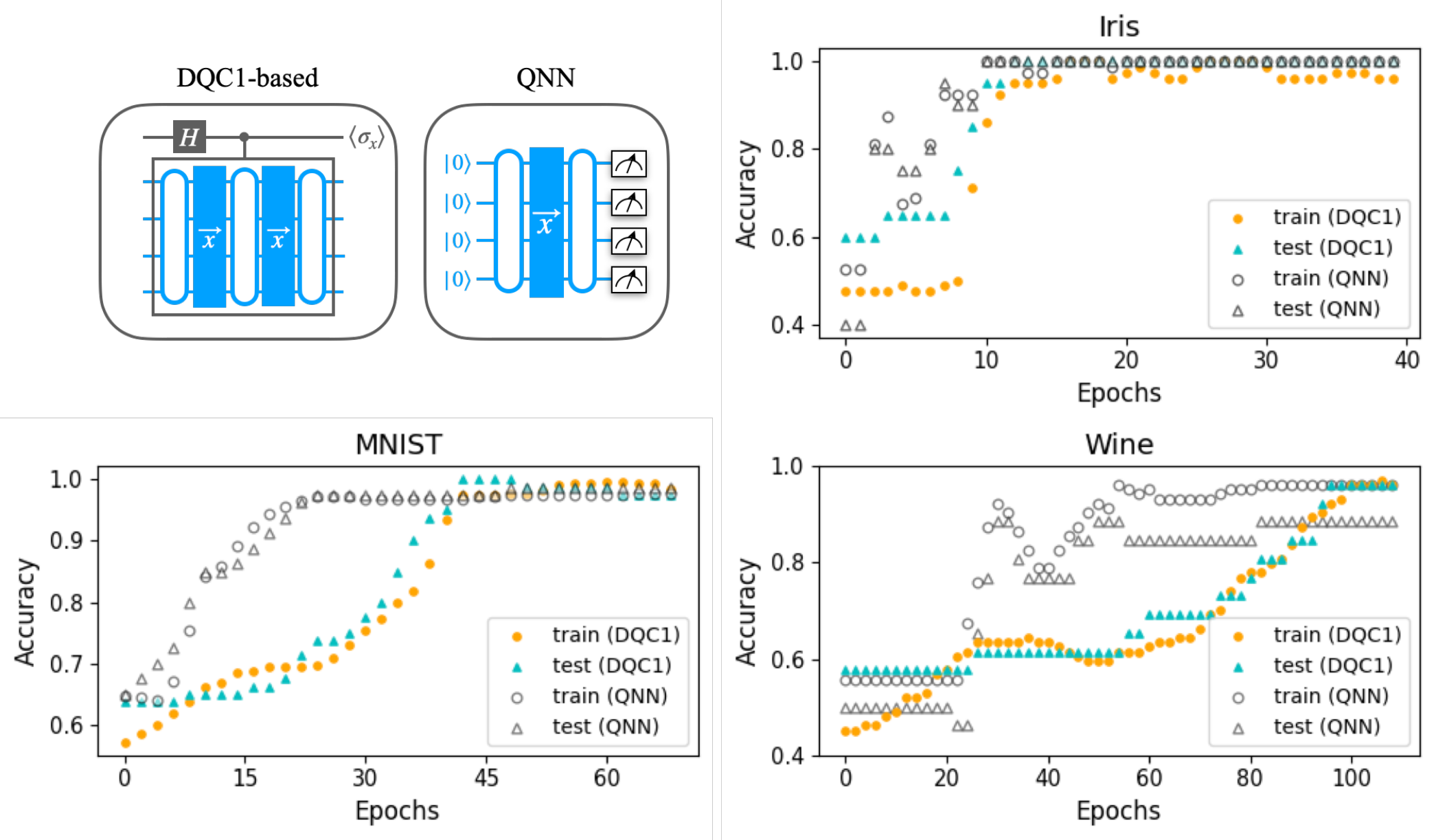}
    \caption{\label{fig:real application} Application of the DQC1-based ML model and the QNN for binary classification on real-world datasets. Each circuit is shown at the top-left. The plots illustrate the accuracy for each dataset, where filled (empty) circles represent the training data accuracy for the DQC1-based (QNN) model, and filled (empty) triangles indicate the test data accuracy for the DQC1-based (QNN) model.}
\end{figure}

Figure~\ref{fig:real application} presents the circuits for the DQC1-based ML model and the QNN model, along with the results of our experiments. 
As in the experiment shown in Fig.~\ref{fig:dqc1_qnn}, we utilized two circuits with the same expressivity, determined by the structure of the data embedding part. Here, however, the ZZ feature map~\cite{Havlicek2019} was chosen as the embedding structure for multivariate inputs instead of single qubit rotation gates, with the following conventional form:
\begin{equation}
U_{\phi}(\boldsymbol{x})
=\left[\exp(i\sum_i\phi_i(\boldsymbol{x})Z_i+i\sum_{i,j}\phi_{i,j}(\boldsymbol{x})Z_iZ_j)
H^{\otimes n}\right]^L\,,
\end{equation}
where $\phi_i(\boldsymbol{x})=\boldsymbol{x}_i$ and $\phi_{i,j}=(\pi-\boldsymbol{x}_i)(\pi-\boldsymbol{x}_j)$.
For the trainable unitary, Ansatz 1 from Fig.~\ref{fig:ml example} was used in both cases.

The results verify the applicability of the DQC1-model and demonstrate its performance at a level comparable to that of a universal quantum model.
For the Iris dataset, both models performed well, achieving nearly 100\% accuracy on both the training and test datasets after around 10 epochs. For the MNIST dataset, both models reached nearly 100\% accuracy on training and test datasets as well. Although the DQC1-based model converged at a slower rate, it ultimately achieved higher accuracy on both training and test datasets compared to the QNN model. For the Wine dataset, the DQC1-based model achieved approximately 96\% accuracy on both the training and test datasets. In contrast, the QNN model reached around 96\% accuracy on the training dataset but only 88\% on the test dataset. While the QNN model tended to converge faster, the DQC1-based ML model ultimately demonstrated higher final test accuracy, consistent with the trend observed for the MNIST dataset.

\section{DQC1 at Finite Temperatures}
\label{sec:finite}

In the conventional DQC1 protocol, the working register is initialized in the maximally mixed state, corresponding to the thermal equilibrium state at infinite temperature. In practice, the quantum system is often provided in the thermal equilibrium state at finite temperature. If the working register of DQC1 is initialized in an arbitrary density matrix $\rho_{w}$, the final $2^{n+1}\times 2^{n+1}$ state of the DQC1 protocol can be expressed as
\begin{equation}
    \rho=\frac{1}{2}\left(I\otimes \rho_w+\alpha\left(|0\rangle\langle 1|\otimes \rho_w U^{\dagger}+|1\rangle\langle 0|\otimes U\rho_w\right)\right).
\end{equation}
Measuring the expectation values of $\sigma_x$ or $\sigma_y$ on the signal qubit yields
\begin{align}
\begin{split}
\langle \sigma_x\rangle&=\frac{\alpha}{2^n}\mathrm{Re}\left(\mathrm{tr}\left(\rho_wU(\boldsymbol{x},\boldsymbol{\theta})\right)\right),\\
\langle \sigma_y\rangle&=\frac{\alpha}{2^n}\mathrm{Im}\left(\mathrm{tr}\left(\rho_wU(\boldsymbol{x},\boldsymbol{\theta})\right)\right).
\end{split}
\end{align}
Therefore, the DQC1-based ML model with $n$ uniformly random bits can produce the following set of functions: 
\begin{equation}\label{eq:fx_finite}
f(\boldsymbol{x},\boldsymbol{\theta}) \in \left\lbrace \frac{1}{2^n}\mathrm{tr}\left(\rho_w U(\boldsymbol{x},\boldsymbol{\theta})\right) : U\in U(2^n), \boldsymbol{\theta}\in \boldsymbol{\Theta} \right\rbrace.
\end{equation}
The form of $\mathrm{tr}\left(\rho_w U(\boldsymbol{x},\boldsymbol{\theta})\right)$ can also be expressed as a Fourier-type sum as in Eq.~\eqref{eq:trace_fourier}, with

\begin{equation}
\exp(\,i\omega(\boldsymbol{x}))
=\exp\left(\,i\sum^{L}_{i=1} (D_{i})_{k_{i}k_{i}}
\right)\,,
\end{equation}
and
\begin{equation}
    c_{\omega}(\boldsymbol{\theta})
=(\rho_w W_{1}(\boldsymbol{\theta}_{1}))_{k_{L}k_{1}}
(W_{2}(\boldsymbol{\theta}_{2}))_{k_{1}k_{2}} \cdots
(W_{L}(\boldsymbol{\theta}_{L}))_{k_{L-1}k_{L}}.
\end{equation}
This demonstrates that varying the temperature of the working register alters only the coefficients of the Fourier series, while the frequency components---and thus the expressivity---remain unchanged.

\section{DQC1 with Multi-qubit Measurement}
\label{sec:multi-measurement}

The DQC1 protocol can be extended to incorporate measurements on the working registers~\cite{PhysRevLett.112.130502}. Despite this extension, the model remains subuniversal, as the number of qubits with non-zero purity is still limited to one. However, this modification provides greater flexibility compared to the original DQC1, as measurements are no longer confined to a single qubit.

The final state of this modified DQC1 protocol is the same as described in Eq.~(\ref{eq:measure}). However, we now assume that the expectation values of observables such as $\sigma_x^{(\mathrm{s})} \otimes M^{(\mathrm{w})}$ and $\sigma_y^{(\mathrm{s})} \otimes M^{(\mathrm{w})}$ can be estimated, where $M^{\dagger} = M$. Here, the superscripts $\mathrm{s}$ and $\mathrm{w}$ indicate that measurements are performed on the signal and working qubits, respectively.

The expectation value of $\sigma_x^{(\mathrm{s})} \otimes M^{(\mathrm{w})}$ can be calculated as:
\begin{align}
    \langle\sigma_x^{(\mathrm{s})} \otimes M^{(\mathrm{w})}\rangle &= \frac{\alpha}{2^{n+1}}\mathrm{tr}\left(\sigma_x |0\rangle\langle 1|\otimes MU^{\dagger}+\sigma_x|1\rangle\langle 0|\otimes MU\right)\nonumber \\
    & =\frac{\alpha}{2^{n}}\mathrm{Re}\left(\mathrm{tr}\left(MU\right)\right).
\end{align}
A similar calculation yields: 
\begin{equation}
    \langle\sigma_x^{(\mathrm{s})} \otimes M^{(\mathrm{w})}\rangle = \frac{\alpha}{2^{n}}\mathrm{Im}\left(\mathrm{tr}\left(MU\right)\right).
\end{equation}
Thus, the modified DQC1 protocol with multi-qubit measurement can generate the following set of functions: 
\begin{equation}\label{eq:fx_finite}
f(\boldsymbol{x},\boldsymbol{\theta}) \in \left\lbrace \frac{1}{2^n}\mathrm{tr}\left(M U(\boldsymbol{x},\boldsymbol{\theta})\right) : U\in U(2^n), M^{\dagger}=M, \boldsymbol{\theta}\in \boldsymbol{\Theta} \right\rbrace.
\end{equation}

Following a similar analysis to that in Sec.~\ref{sec:finite}, we find that multi-qubit measurement does not enhance expressivity unless the measurement operator $M$ is a function of the input data $\boldsymbol{x}$. For example, the cardinality of the frequency spectrum can increase to $2^{n(L+2)}$ if the input data $\boldsymbol{x}$ is encoded into the measurement operator as $M(\boldsymbol{x})=U(\boldsymbol{x})^{\dagger}MU(\boldsymbol{x})$, where $U(\boldsymbol{x})$ is a unitary circuit parameterized by $\boldsymbol{x}$.

\section{Conclusions and Discussions}
\label{sec:conc}

In this work, we introduced the DQC1-based ML model and analyzed its expressive power. Our theoretical analysis and numerical simulations demonstrate that this subuniversal model of quantum computation is equally capable as the universal model in generating the Fourier basis. Since the DQC1 protocol is well-suited for ensemble quantum information processors such as those with spin ensembles and magnetic resonance, our findings broadens the range of feasible quantum computing hardware platforms for QML. Although DQC1 traditionally relies on having $n$ qubits in the maximally mixed state, we have shown that its expressive power remains unchanged even when using states in thermal equilibrium at finite temperatures. This enhances the practicality of subuniversal quantum computing models for real-world applications.

Since only one qubit requires non-zero polarization and measurement, the DQC1-based ML offer advantages over universal quantum circuits regarding the state preparation and measurement errors. Specifically, efforts to minimize these errors can be concentrated on a single qubit. Moreover, when building hardware for DQC1, only one qubit needs to be coupled with a measurement device, which simplifies the overall architecture. On the other hand, DQC1 circuits place more stringent demands on control fidelity and coherence time, as all unitary operations must be transformed into controlled operations. Assuming that an $n$-qubit unitary operator $U$ is decomposed into single-qubit and two-qubit entangling gates (e.g., CNOT), converting $U$ into a controlled-$U$ with a single control qubit increases the circuit depth by a constant factor~\cite{Nielsen:2011:QCQ:1972505,PhysRevA.106.042602}. Consequently, the control fidelity and coherence time required for the DQC1 protocol should exceed those needed in universal quantum computation by at least a constant factor. Errors occurring on the working register during the controlled unitary operation can propagate to the signal qubit and effectively become part of the initialization or measurement errors of the signal qubit. The initialization error can be modeled as a stochastic Pauli channel (with twirling applied if necessary~\cite{PhysRevA.94.052325,PhysRevA.106.012439}), which prepares the signal qubit in the state $(I+\alpha'\sigma_z)/2$ with $\alpha'<\alpha$. This reduction in polarization results in an increase in the number of repetitions of the DQC1 circuit by a factor of $(\alpha/\alpha')^2$. Similarly, measurement errors reduce the expectation values in Eq.~(\ref{eq:observables}), also necessitating additional repetitions. Incorporating existing quantum error mitigation techniques~\cite{endo2018practical,temme2017error,9226505,Kurita2023synergeticquantum,Maciejewski2020mitigationofreadout,9142431,kim2022quantum,PhysRevApplied.17.014024,Lee_2023} or developing new methods specific to DQC1 could improve the utility of NISQ devices for DQC1, making them interesting directions for future work.

A major challenge in both QNNs~\cite{10.1038/s41467-018-07090-4,larocca2024review} and quantum kernel methods~\cite{Thanasilp2024} is the exponential concentration phenomenon. In the context of QNNs, this phenomenon is also referred to as barren plateaus, as it can be understood as the parameter optimization landscape becoming exponentially flat. The DQC1-based machine learning model also suffers from exponential concentration, but from a different source. 

To examine this issue in detail, let us first consider the quantity $\mathrm{tr}(U)$. In an $n$-qubit system, the trace of a unitary matrix $U$ is a sum of its diagonal elements, $\mathrm{tr}(U)=\sum^{2^n}_{i=1}U_{ii}$, where each diagonal entry must have an absolute value of at most 1, $|U_{ii}|\leq 1$. The only relationship all entries must adhere to is $\sum^{2^n}_{k=1}|U_{ik}|^2=\sum^{2^n}_{k=1}|U_{ki}|^2=1$ for all $i=1,\cdots,2^n$ derived from $U^{\dagger}U=UU^{\dagger}=I_{n}$ and the diagonal entries do not have specific interdependencies. Under the condition of independent diagonal elements with each size at most 1, Hoeffding's inequality provides an upper bound on the probability, 
\begin{align}
\begin{split}\label{eq: devia}
        &\mathrm{Pr}_U [|\mathrm{tr}(U)-\mathbb{E}_U[\mathrm{tr}(U)]|\geq t]
        \leq 4 \exp(-\frac{t^2}{2^{n+1}}),   
\end{split}    
\end{align}
for all $U$. 
These conditions do not exhibit exponential concentration with respect to the number of qubits $n$. For a fixed $n$, the probability that $\mathrm{tr}(U)$ deviates from its expected value by $t$ decreases exponentially with $t^2$. This implies that it becomes exponentially unlikely to obtain $\mathrm{tr}(U)$ differing significantly from its expectation value. However, this exponential concentration effect diminishes as the number of qubits $n$ increases. Thus, the fundamental issue of exponential concentration does not arise when computing $\mathrm{tr}(U)$. On the other hand, the ML model in DQC1 incorporates the factor $2^{-n}$. In this case, the inequality becomes
\begin{equation}\label{eq: devia2}
        \mathrm{Pr}_U [|f(\boldsymbol{x},\boldsymbol{\theta})-\mathbb{E}_U[f(\boldsymbol{x},\boldsymbol{\theta})]|\geq t]
        \leq 4 \exp(-2^{n-1}t^2).
\end{equation}
This demonstrates that a DQC1 model is \textit{exponentially concentrated} with respect to the number of qubits.

We can reach the same conclusion that the exponential concentration in a DQC1 model is not because of $\mathrm{tr}(U)$, but instead due to the factor $2^{-n}$, through an alternative approach.
There is a proof of convergence in distribution for $\mathrm{tr}(U)$ that if $U$ follows a Haar distribution, $\mathrm{tr}(U)$ converges to a standard complex normal $\mathcal{CN}(0,1)$ as $n\rightarrow\infty$~\cite{petz2003asymptoticslargehaardistributed}. It is equivalent that $\mathrm{Re}(\mathrm{tr}(U))$ and $\mathrm{Im}(\mathrm{tr}(U))$ are independent and both follow standard normal distribution, $\mathcal{N}(0,1/2)$, by definition. Although this result holds under the specific condition of Haar-uniform randomness, it demonstrates that there is no explicit effect of concentration over $n$. However, by multipyling a factor of $2^{-n}$, we obtain
\begin{equation}\label{eq: variance}
\mathrm{Var}_U[f(\boldsymbol{x},\boldsymbol{\theta})]
    =\frac{C(n)}{4^n}
\end{equation}
where $C(n)$ is a value satisfying $\lim_{n\rightarrow\infty}C(n)=(1+i)/2$. Therefore, if $\boldsymbol{\theta}$ is sampled such that $U(\boldsymbol{x},\boldsymbol{\theta})$ is Haar random, then the distribution of $f(\boldsymbol{x},\boldsymbol{\theta})$ concentrates to its expectation value, which is zero, exponentially fast in $n$. More explicitly, according to the Chebyshev inequality, the probability of $f(\boldsymbol{x},\boldsymbol{\theta})$ deviating from its mean of zero is bounded as
\begin{equation}
        \mathrm{Pr}_U [|f(\boldsymbol{x},\boldsymbol{\theta})|\geq \varepsilon]
        \leq \frac{\mathrm{Var}_U[f(\boldsymbol{x},\boldsymbol{\theta})]}{\varepsilon^2}
\end{equation}
for any real number $\varepsilon>0$.
The variance is exponentially small in $n$, as shown in Eq.~(\ref{eq: variance}), when $U$ is Haar random. Hence, a DQC1 model experiences exponential concentration in this case.

While exponential concentration is generally perceived as an obstacle, a potential future direction is to explore the advantages of DQC1 in building highly sensitive ML models for detecting small fluctuations around zero and predicting functions with extremely small values, and to find practical applications for these models. Another important future work is to investigate the trade-off between the distance of the distribution that $U(\boldsymbol{x},\boldsymbol{\theta})$ follows from the Haar distribution (or its distance from being a 2-design) and the concentration phenomenon, in a similar spirit to Ref.~\cite{holmes_2022}. Identifying other resources that lead to exponential concentration in DQC1 models and comparing the rates of concentration to the universal model also suggest further exploration. Exploring the effect of entanglement between the signal qubit and the working register on the exponential concentration phenomenon is of particular interest. In this context, the signal qubit and the working register can be considered as the visible and hidden units in a QNN, where entanglement between them is known to induce barren plateaus~\cite{PRXQuantum.2.040316}. However, the DQC1 circuit cannot generate entanglement between them when the purity ($\alpha$) is below a certain threshold~\cite{PhysRevA.72.042316,PhysRevA.95.022330}. Therefore, it remains an intriguing open question whether a quantum advantage can be achieved via the DQC1 model while avoiding the entanglement-induced exponential concentration phenomenon. 

Investigating solutions to the general problem of exponential concentration remains a crucial area for future research. In particular, adapting techniques from variational quantum algorithms---such as variable structure ansatz~\cite{AdaptVQE,QMI_semiagnostic,AdaptVQE-2,Wada_2024}, initialization strategies~\cite{Grant2019initialization,10.5555/3600270.3601622,wang2023trainability}, pre-training methods~\cite{PRXQuantum.2.020329,PhysRevA.106.042433,synergistic}, layer-wise training~\cite{layerwise_learning}, entanglement regularization~\cite{PhysRevResearch.3.033090}, and advanced optimization methods~\cite{PhysRevResearch.6.023069}---is a promising path forward. Developing new techniques specifically tailored to DQC1 models is another important direction to consider.

The connection to quantum signal processing (QSP), noted in Section~\ref{sec:expressivity}, also points to an intriguing direction for future work. While the quantum circuit structures investigated here primarily focus on learning from classical data inputs, extending this framework to quantum data applications by encoding matrix inputs, such as Hamiltonians, through block encoding---as in quantum singular value transformation (QSVT)~\cite{Gily_n_2019}---warrants further investigation.

\section*{Data Availability}
The data and software that support the findings of this study can be found in the following repository: \url{https://github.com/qDNA-yonsei/DQC1-QML}.

\section*{Acknowledgments}
This work was supported by Institute of Information \& communications Technology Planning \& evaluation (IITP) grant funded by the Korea government (No. 2019-0-00003, Research and Development of Core technologies for Programming, Running, Implementing and Validating of Fault-Tolerant Quantum Computing System), the Yonsei University Research Fund of 2024 (2024-22-0147), the National Research Foundation of Korea (Grant No. 2023M3K5A1094805, 2023M3K5A1094813), the KIST Institutional Program (2E32941-24-008).

\appendix
\section{Optimizer and Learning Rate Selection}\label{Append: A}

\begin{figure}[t]
    \centering
\hspace{-1cm}    
\includegraphics[width=0.7\columnwidth]{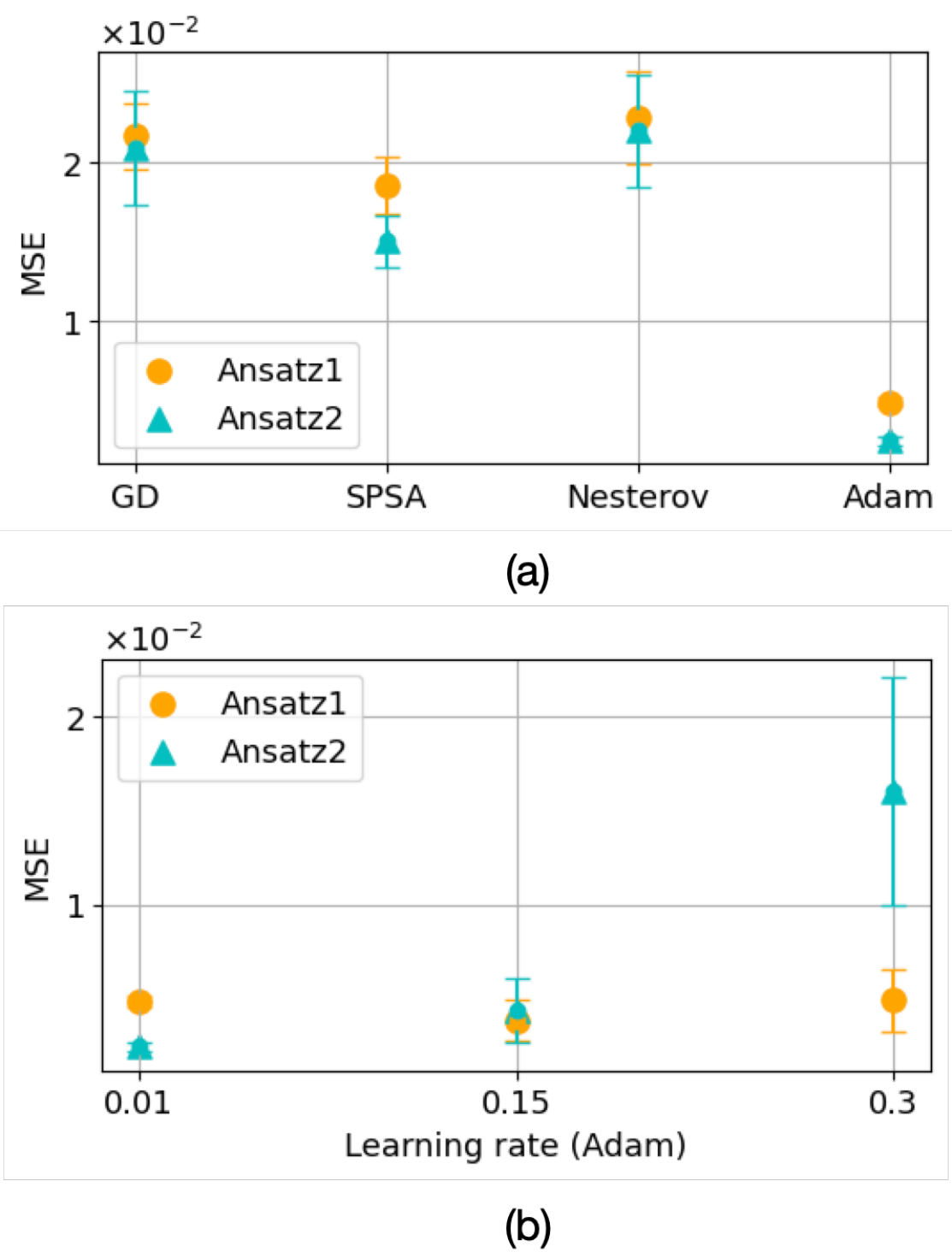}
    \caption{\label{fig:optimizer_lr} (a) Comparative analysis of different optimizers, showing the final MSE for each optimizer after training. The markers indicate the mean, and error bars represent standard deviation, calculated from five repetitions with randomly initialized parameters. All experiments were conducted to learn the target function $g_{3}(x)=\sum_{k=-4}^{4}c_{k}\exp(ikx)$ using a $4$-qubit circuit with $|\Omega|=9$, which is the fourth circuit from the left in Fig.~\ref{fig:ml example}. 
    The input data and the two ans\"{a}tze used are the same as those in Fig.~\ref{fig:ml example}. Parameter updates were iterated 200 times with the learning rate set to the default value of each optimizer provided in Pennylane~\cite{bergholm2020pennylane}. The Adam optimizer achieved the best performance. (b) Comparative analysis of the learning rate for the Adam optimizer, displaying the final MSE obtained by varying the learning rate of the Adam optimizer under the same conditions as in (a).}
\end{figure}

In this appendix, we present a comparative analysis of several optimizers based on their final MSE after training. Additionally, we explore the impact of different learning rates on the performance of the best-performing optimizer. All of these results are illustrated in Fig.~\ref{fig:optimizer_lr}.

The analysis uses the 4-qubit circuit with $|\Omega|=9$, which requires the largest number of parameters for the target function $g_3(x)=\sum_{k=-4}^{4}c_{k}\exp(ikx)$ in Fig.~\ref{fig:ml example}. We utilize a total of four optimizers from PennyLane~\cite{bergholm2020pennylane}: gradient descent, simultaneous perturbation stochastic approximation (SPSA)~\cite{119632}, Nesterov momentum~\cite{Nesterov1983AMF}, and Adam~\cite{kingma2017adam}. The results show that the Adam optimizer achieves the lowest MSE, making it the optimizer of choice for all experiments in this work. 

Further experiments are conducted to identify the optimal learning rate for the Adam optimizer. The results suggest that learning rates of 0.01 and 0.15 yield the lowest MSE for each ansatz, with a learning rate of 0.15 producing consistently low values across both ans\"{a}tze; hence, this rate is adopted for all experiments.

\clearpage

\end{document}